\documentclass[a4paper]{spie}  
\usepackage{amsmath,amsfonts,amssymb}
\usepackage{graphicx}
\usepackage[colorlinks=true, allcolors=blue]{hyperref}
\usepackage{colortbl}
\usepackage{subcaption}

\title{CHARA/SPICA: a 6-telescope visible instrument for the CHARA Array}

\author[a]{Denis Mourard}
\author[a]{Philippe Berio}
\author[a]{Cyril Pannetier}
\author[a]{Nicolas Nardetto}
\author[a]{Fatme Allouche}
\author[a]{Christophe Bailet}
\author[a]{Julien Dejonghe}
\author[a]{Pierre Geneslay}
\author[a]{Estelle Jacqmart}
\author[a]{Stéphane Lagarde}
\author[a]{Daniel Lecron}
\author[a]{Frédéric Morand}
\author[a]{Sylvain Rousseau}
\author[a]{David Salabert}
\author[a]{Alain Spang}
\author[b]{Simon Albrecht}
\author[c]{Narsireddy Anugu}
\author[d]{Laurent Bourges}
\author[c]{Theo A. ten Brummelaar}
\author[a]{Orlagh Creevey}
\author[e]{Sebastien Deheuvels}
\author[a]{Armando Domiciano de Souza}
\author[c]{Doug Gies}
\author[a]{Roxanne Ligi}
\author[d]{Guillaume Mella}
\author[f]{Karine Perraut}
\author[c]{Gail Schaefer}
\author[g]{Markus Wittkowski}

\affil[a]{Laboratoire Lagrange, Universit\'e C\^ote d’Azur, Observatoire de la C\^ote d’Azur, CNRS, Parc Valrose, 06108 Nice, France}
\affil[b]{Aarhus Univ. (Denmark)}
\affil[c]{CHARA Array, Georgia State University, Atlanta, GA 30302, USA}
\affil[D]{Univ. Grenoble Alpes, CNRS, IRD, INRAE, Météo France, OSUG, 38000 Grenoble, France}
\affil[e]{Institut de Recherche en Astrophysique et Planétologie}
\affil[f]{Univ. Grenoble Alpes, CNRS, IPAG, 38000 Grenoble, France}
\affil[g]{European Southern Observatory (Germany)}

\begin{document} 
\maketitle

\begin{abstract}
With a possible angular resolution down to 0.1-0.2 millisecond of arc using the 330~m baselines and the access
to the 600-900 nm spectral domain, the CHARA Array is ideally configured for focusing on precise and accurate fundamental parameters of stars. CHARA/SPICA (Stellar Parameters and Images with a Cophased Array)
aims at performing a large survey of stars all over the Hertzsprung-Russell diagram. This survey will also study the effects of the different kinds of variability and surface structure on the reliability of the extracted fundamental parameters. New surface-brightness-colour relations will be extracted from this survey, for general purposes on distance determination and the characterization of faint stars. SPICA
is made of a visible 6T fibered instrument and of a near-infrared fringe sensor. In this paper, we detail the science program
and the main characteristics of SPICA-VIS. We present finally the initial performance obtained during the
commissioning.
\end{abstract}


\section{The Science perspective of CHARA/SPICA}
\label{sec:science}
Precise fundamental stellar parameters are the primary data required for an in-depth understanding of stellar evolution, interiors, and environments. With the progress of stellar physics and the prospects of ground facilities or space missions, it is critical to improve the accuracy, and quantity of such data. The development of exoplanet and asteroseismology domains is demanding direct data to eliminate any bias in the parameters (Radius, Effective Temperature, gravity...)\cite{gent21}. Besides, the current tension on the Hubble constant is motivating new and precise direct determination of distance of the primary candles of the cosmic scales\cite{riess16, riess19}. Many methods like asteroseismology, photometric transits of exoplanets, radial velocities, or Surface Brightness Colour relations (SBCR) are linked to the stellar radius. Usually estimated through models, its determination by coupling an optical interferometric measurement of the angular diameter and, for example, a \textit{Gaia} parallax, is the best way to avoid any model dependence. Furthermore, characterizing any activity (limb darkening, rotation, winds, or binarity) is also mandatory, both for bias removal and for the required progress on stellar physics. 

With this general framework in mind, we developed the idea of an Interferometric Survey of Stellar Parameters (ISSP) with about 1000 stars with a declination $\delta>-30^{\circ}$ and a magnitude in the R band brighter than 8. The survey is built around four main scientific questions:

\subsection{Radius and Effective Temperature of a large sample of exoplanet host stars ($\approx$~50 stars)}
For this program we focus on known systems and especially transiting ones. In the exoplanet field, while efforts are focused on planet detection and characterization, knowledge of the star (radius, age, mass, activity) is mandatory to determine the parameters of the planet and understand the formation and evolution of exoplanetary systems \cite{lebreton14}. CoRoT and Kepler have fulfilled their mission and brought an impressive number of transiting planet candidates, but the follow-up is difficult because of the faint magnitude of the stars. The next step is to go further and to detect less massive planets, possibly in the habitable zone, or to study in detail the atmospheric properties of larger planets. This can only be done on brighter targets and future missions are focusing precisely on these kinds of stars: TESS, CHEOPS, PLATO, and ARIEL. Precise determinations of radii, masses and ages of exoplanets have been already demonstrated but on a very small number of targets \cite{ligi12,ligi16}. We plan to combine directly measured angular diameters with parallaxes and bolometric fluxes to extract the radius and the effective temperature. These quantities will also be used to constrain the evolutionary status of the systems and to derive their age. These estimations will permit us to improve the star-planet relationships, and to measure and unbias the parameters of the planet. These topics are built on previous works but one of the most ambitious objectives of ISSP is to reach a sample large enough for statistical considerations, and with the precision required for the future PLATO targets, the overall goal being a unique contribution to the question of the existence of other habitable planets.

\subsection{Observation of a large and homogeneous sample of asteroseismic and interferometric targets ($\approx$~300 stars)}
This sample is focused on bright ($V<8$) F5-K7 spectral type sources (class V/IV and III), according to the PLATO mission selection. In recent years, photometric missions (WIRE, MOST, CoRoT, and Kepler) have been providing very long (weeks to years) nearly uninterrupted time series of pulsations in stars across the HR diagram, and for thousands of giant stars. While seismic data of the flux variations caused by pulsations have become available for a good sample of stars, independent measures of many radii have not been possible due to the small angular diameter of the stars and their faint magnitude. The limitation in exploiting this potential can be overcome by studying their nearby and bright counterparts, for which both angular diameters and seismic data (among other things) are accessible. This program will permit, among others, the absolute and direct calibration of the scaling relations used in asteroseismology. These relations are classically used to extract the radius and the mass from the frequency measurements. One of the main questions will be about the universality of these scaling relations or their fine structure (metallicity, luminosity class). Coupling interferometry and asteroseismology has brought critical advances to stellar modelling \cite{bigot11,perraut13} and these programs should be extended to larger samples over a wider region of the HR diagram.

\subsection{Calibration of new SBCR ($\approx$~300 stars)}
The initial ISSP catalogue is constructed to allow the identification of stars presenting a very low level of activity. These stars will be used to calibrate new SBCR without the intrinsic bias that any activity could introduce. Our goal with the large and carefully selected sample of ISSP is to reach a precision and accuracy of 1-2\% for the improvement of distance determination of eclipsing binaries in LMC \cite{gp13, gallenne18, gp19}, SMC \cite{graczyk20}, M31 and M33.  
and to the faint targets of PLATO \cite{gent21}. The size and diversity of the survey should, in principle, permit us to differentiate the SBCR as a function of luminosity class and metallicity \cite{salsi20,salsi21,salsi22}. Direct imaging of stellar surfaces by long baseline optical interferometry has also started to bring the required information on activity markers. For signal to noise reasons, it is however limited to bright and stars with large enough angular diameters. In this domain, our ambition is primarily oriented toward the identification and characterization of the departures, in the interferometric data, from the simple uniform disk model: limb darkening, fast rotation, multiplicity, spots, or environments. These direct constraints are mandatory to correctly disentangle stellar noise in planet detection and characterization, to improve the accuracy and precision in the SBCR, and finally to improve the stellar models. Complementary homogeneous photometric (in particular in the infrared) and spectroscopic data will be mandatory. The impact of these new SBCR for the question of distance estimations in the Universe is critical. This part of the ISSP survey is closely coordinated with the Araucaria Project \footnote[1]{Araucaria project:~\url{https://araucaria.camk.edu.pl/}}, whose purpose is to provide an improved local calibration of the extragalactic distance scale out to distances of a few megaparsecs~(\cite{gieren05_messenger}). 

\subsection{Markers of activity through intensity profile and surface structure ($\approx$~350 stars)}
This sample is dedicated to the direct measurement of the intensity profile, either by empirical limb-darkening laws, or directly by MARCS or 3D Stagger intensity profile adjustments. This will improve the elimination of bias or uncertainties in the transit analysis for systems where both photometric and interferometric measurements are possible. This unique and large survey will also permit us to sample a wide variety of spectral types and luminosity classes and to establish actual limb darkening profiles to be used for faint stars. This study of limb darkening will be extended to more complex activity markers like fast rotation, and multiplicity or circumstellar environment. These characterizations by imaging will permit us to disentangle the effects of these physical processes in the determination of the angular diameter with a direct impact in the calibration and the future use of the SBCR relations \cite{challouf15}. We are also expecting that these measures will directly aid progress towards a better understanding of the physics of stars and their atmospheres.

\begin{figure} [htbp]
\begin{center}
\includegraphics[width=0.7\textwidth]{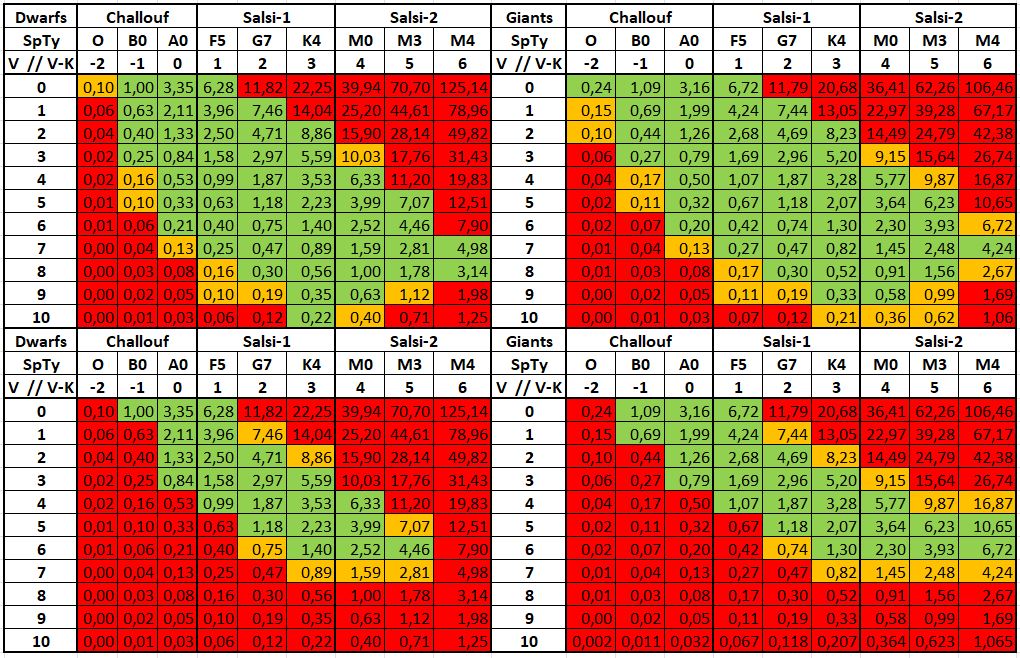}
\caption{Possible domains (in green) of observations of CHARA/SPICA for dwarfs (left column) and giants (right columns). The red boxes are out of reach whereas the orange ones represent the current limits. The value in each cell is an estimation of the angular diameter based on various SBCR relations \cite{salsi20,salsi21,challouf14}. The criterion is based on a 1\% precision estimation of the stellar radius with an a-priori hypothesis on the limb-darkening (upper row) or without a-priori (lower row). In this last case, both the angular diameter and the intensity profile will be extracted.}
\label{fig:science} 
\end{center}
\end{figure} 

\subsection{Additional programs and expected performance}
To validate the possibility of such a large survey with a new 6-telescope visible instrument on the CHARA Array, we developed during the last years a technological program based on single mode fibers, EMCCD detectors, and optimization of the injection into single mode fibers after a partial adaptive optics correction\cite{Martinod2018}. This program has permitted us to validate a certain number of hypotheses on the SPICA performance, and we built a signal to noise ratio estimator derived from the FRIEND experiment\cite{Martinod2018} and adapted to the SPICA design. Based on this estimator and on numerical simulations performed with ASPRO2 and LitPro software\footnote[2]{Available at~\url{http://www.jmmc.fr}} we predicted the possible domain of operation of SPICA for dwarfs and giants, as a function of the V magnitude and of the (V-K) color index. The results are presented in Figure~\ref{fig:science}.

Even though the survey presented above constitutes just the core of the science objective of CHARA/SPICA\cite{mourard2017,spie2018}, the expected instrumental capabilities of the instrument will enable many other programs \footnote[3]{CHARA/SPICA website:~\url{https://lagrange.oca.eu/fr/spica-project-overview}} . Initially designed for high sensitivity and high precision measurements, the instrument offers three different spectral resolution (200, 4000, 13000) with spectral bands of 300, 90 and 30 nm, respectively.\\

The SPICA-VIS visible instrument and its science program have been designed to be accompanied by a fringe tracking capability with SPICA-FT (see Pannetier et al. this conference). To avoid long and expensive developments, it has been decided to develop, first, a dedicated integrated-optics combiner that is installed in front of the MIRC-X spectrograph and, second, to develop an optical path difference controller that enables a servo-loop for fringe stabilization on the main delay lines of CHARA. This development has opened the possibility of the simultaneous operation of SPICA and MIRC-X+MYSTIC\cite{anugu2020} instrument. When this operational model is validated, simultaneous observations in R, J, H, and K band will be permitted.

\section{Design of the SPICA-VIS instrument}
\subsection{SPICA spectrograph}
\begin{figure} [htbp]
\begin{center}
\includegraphics[width=0.6\textwidth]{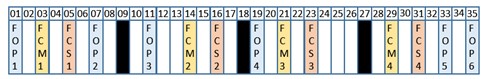}
\caption{Presentation of the arrangement of the six main SPICA fibers (FOP1..FOP6) in the V-groove at the entrance of the spectrograph. 4 multimode fibers (FCM1..FCM4) are also used for the spectral sources, and 4 spare single mode fibers (FCS1..FCS4) are also used for alignment purposes.}
\label{fig:vgroove} 
\end{center}
\end{figure}

\begin{figure} [htbp]
\begin{center}
\includegraphics[width=0.7\textwidth]{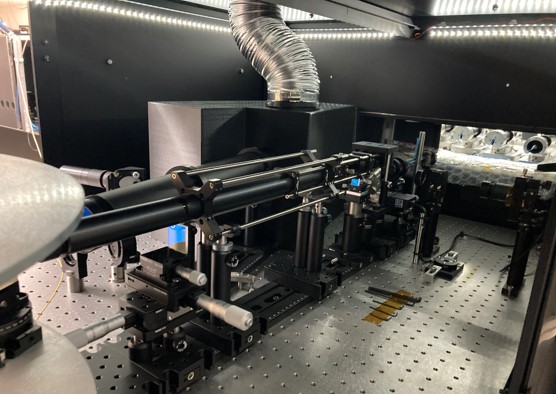}
\caption{General view of the SPICA-VIS spectrograph. The dispersion mechanism is on the left, the light arriving from the right  (the injection modules and the fibers in blue can be seen) through the optomechanical setup holding the anamorphosis system for the interferometric channel and the optics for the transport of the photometric channels. After dispersion, the light is sent by reflection to the detector (with its cooling system) through an optical chamber.}
\label{fig:spectro} 
\end{center}
\end{figure}

\begin{figure}[htbp]
\centering
\includegraphics[width=0.9\textwidth]{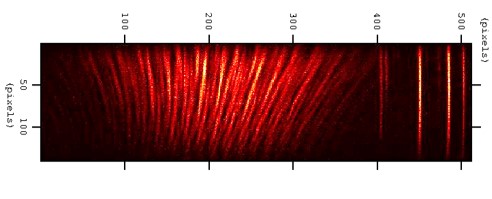}
\caption{An image of the detector in low resolution mode, with the six dispersed photometric channels (from right to left) and the interferometric channel where the 6 beams interfere. The vertical direction is the wavelength from blue (down) to red (top).}
\label{fig:fringes}
\end{figure}

The SPICA-VIS instrument is a fiber-fed spectrograph based on an EMCCD ANDOR Ixon888 detector (1024x1024 pixels of 13$\mu$m). The beams from the six CHARA telescopes are injected into single mode fibers, 1.2m long permitting an efficient spatial filtering. The six main fibers are linearly arranged in a V-groove in a non-redundant scheme, as shown in Fig.~\ref{fig:vgroove}. 
At the output of the Vgroove, 10\% of the light is reflected towards a photometric path  and 90\% transmitted for the interferometric path. The latter forms the dispersed fringes in an image plane on the detector, after an anamorphosis achieving a minimum of 3~pixels sampling on the longest internal baseline. The photometric path reimages and disperses the fiber's output next to the interferogram on the detector for measuring the six photometric channels.
The two paths use the same dispersive component so have the same spectral resolution. To achieve the science goals defined in Sect.~\ref{sec:science}, the spectrograph is composed of a low resolution prism, two gratings (medium and high resolution), and a mirror for a non-dispersed mode used for alignment purpose. The spectrograph is presented in Fig.~\ref{fig:spectro}. Fig.~\ref{fig:fringes} presents an image on the detector in low resolution with the interferometric anamorphosed dispersed channel and the six dispersed photometric channels.

\subsection{SPICA injection table}
To correctly feed the V-groove and the spectrograph with the six CHARA beams, a number of operations on each individual beam is necessary to optimize the level of injected light in the single mode fibers and to maximize the signal to noise ratio of the instrument. This is the role of the injection table. 

\begin{figure}
\centering
\includegraphics[width=0.8\textwidth]{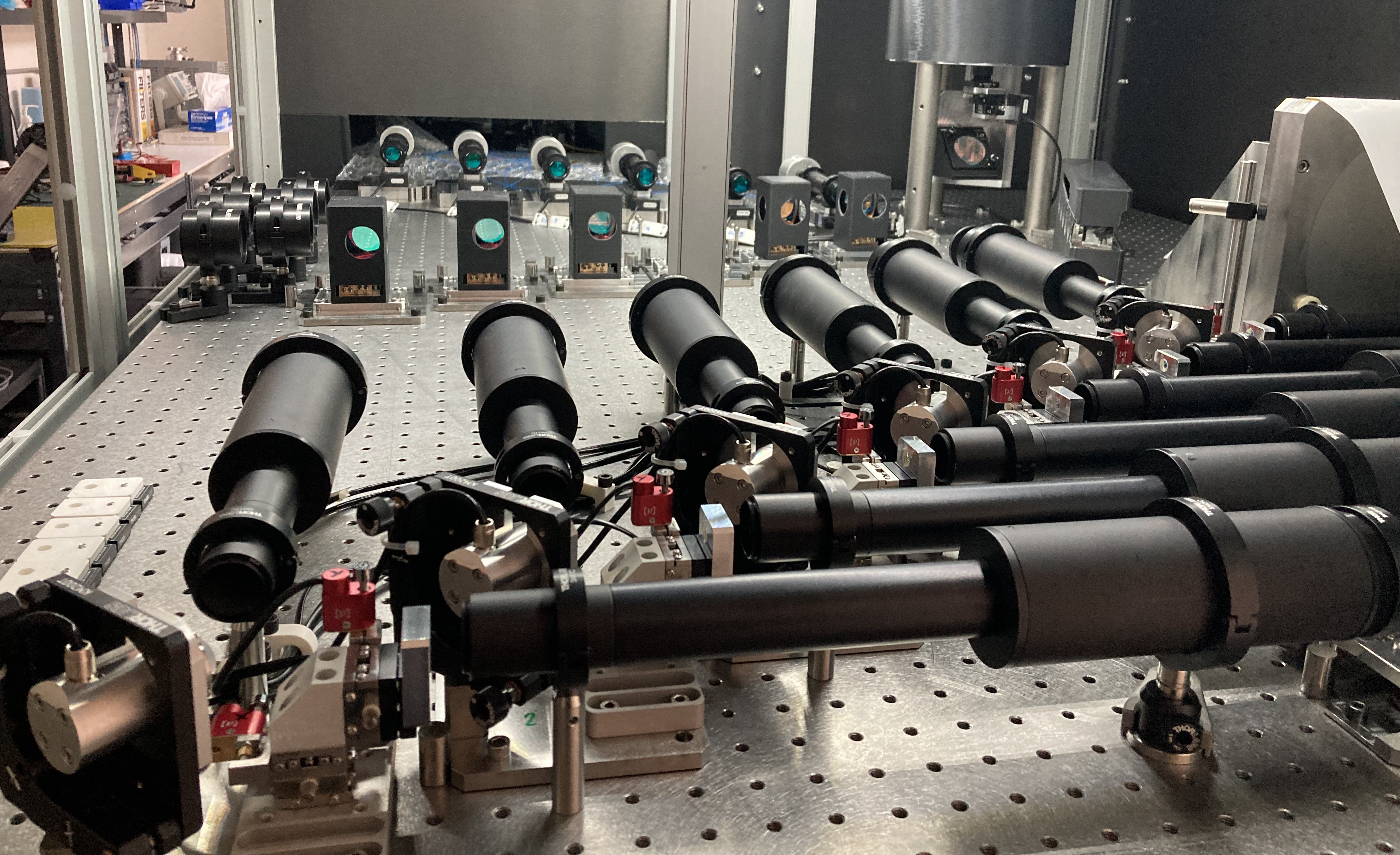}
\caption{The SPICA-VIS injection table. The CHARA beams, arriving from the right on the image, are controlled in orientation and position to correctly feed the injection modules seen on top/left of the image. The control detector is seen on the upper-right part, whereas the beamsplitters and retroreflectors are seen in the middle and left part of the image. In the lower-left corner, one can see the fast tip/tilt mirror (PI company) of the beam 1, receiving the pupil plane reimaged by the field lens that is placed (with its motorizations) in the image plane at the exit of the imaging tube seen on the right. After the tip/tilt mirror, the second collimating tubes send the collimated beams to the injection modules and to the control detector. }
\label{fig:injection} 
\end{figure}

\begin{figure} [htbp]
\begin{center}
\includegraphics[width=0.45\textwidth]{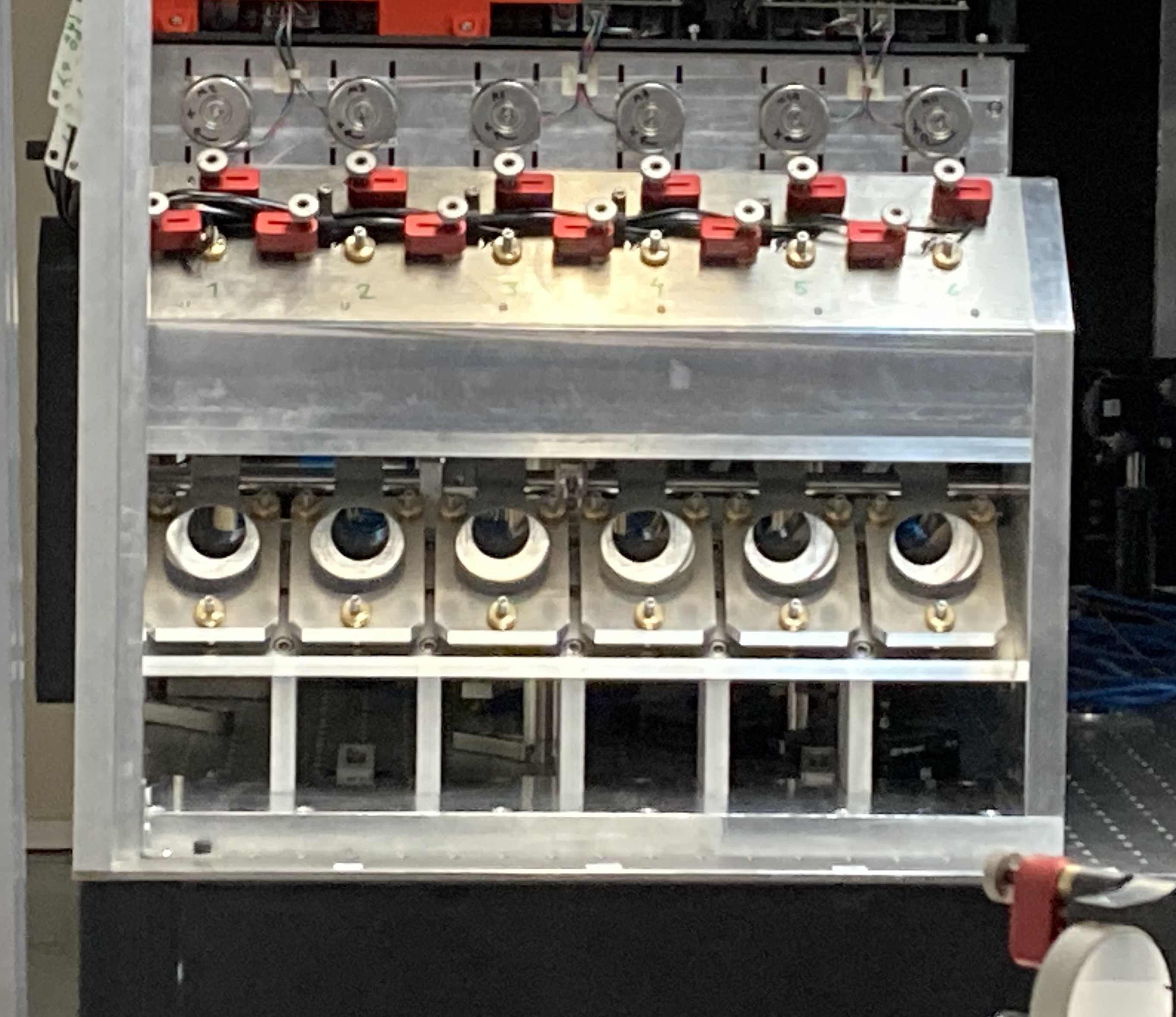}
\hspace{1 cm}
\includegraphics[width=0.3\textwidth]{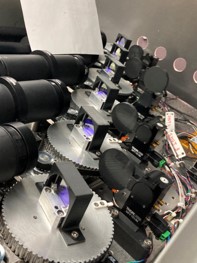}
\end{center}
\caption{In this figure, we see the movable periscope (the 6 bottom mirrors are motorized as a block to permit to the CHARA reference laser to find its way). The 6 upper mirrors cannot be seen but are motorized with piezo actuators (in red). Above these actuators, one can see 6 motors permitting the rotation of the first prism of each ADC, the second one being on the opposite side of the module.}
\label{fig:peri} 
\end{figure}

First of all, a 6-way periscope device (see Fig.~\ref{fig:peri}, left) permit to send the beams to the SPICA tables. The six collimated beams arriving from CHARA have to be controlled both in image and pupil planes (see Fig.~\ref{fig:injection}). Automatic alignment systems are based on images of the stars or of the pupils of the telescopes, recorded on an IXON 897 Andor detector thanks to 10/90 beamsplitters. The control of the position of the images is done on the upper motorized mirrors of the periscope whereas the control of the position of the pupils is done through motorized field lenses installed in an intermediate image plane.

For the pupils a third longitudinal motorization permits a better conjugation of the pupil, projected after the image plane on a fast tip/tilt mirror. This fast tip/tilt mirror is controlled by the real-time analysis of the images on the detector and aims at perfectly stabilizing the corrected image at the entrance of the single mode fibers.

To perfectly feed the V-groove and thus the spectrograph, reference positions on the detector are regularly taken through the fiber-back-illumination device that sends laser light back into the fibers. These retro-beams are directed by the beamsplitters and sent back to the detector thanks to retro-reflectors. 

With the injection in single mode fibers, it is important to compensate for the potential differential birefringence of the fibers. This is done through a plate of polarized crystal that can be smoothly tilted, permitting us to change the polarized delays after the plate in order to match the default of a pair of fibers\cite{Lazareff2012, Martinod2018}. Each beam is equipped with this device, one being considered as reference and which is not rotating (see Fig.~\ref{fig:peri}, right).

Finally, because SPICA is being operated over a spectral band of 300nm in the visible, it is necessary to compensate for the atmospheric refraction for optimizing the injection of light over this large spectral domain. Each beam is thus equipped with an atmospheric dispersion compensator made of two rotating prisms to generate a prism of variable angle. Both prisms rotate in order to follow the field rotation of the CHARA telescopes (see Fig.~\ref{fig:peri}, left).

During the development of the SPICA project we also decided to improve the longitudinal dispersion compensation of CHARA to optimize the interferometer as a whole, to enable routine operation of the different instruments from R to JHK bands. This work has been already published\cite{Pannetier2021}.

\section{The SPICA-VIS data flow}
\begin{figure} [htbp]
\begin{center}
\includegraphics[width=0.8\textwidth]{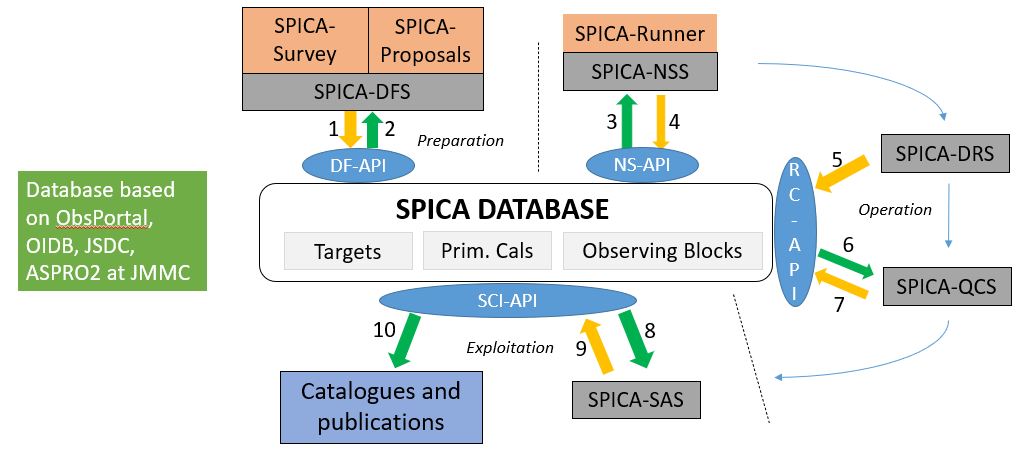}
\end{center}
\caption{General presentation of the SPICA data flow, based on a central database interfaced to various tools for a global and efficient management of the SPICA observations.}
\label{fig:spicadb} 
\end{figure}

In parallel with the automation of the instrument and of its optimisation, we have developed a detailed model of the data flow for the management of the survey over many semesters of observation. This concerns successively the preparation of the observing programs, the night scheduling, the data reduction software and the quality control software, and the data archiving and referencing in the different databases (see Fig.~\ref{fig:spicadb}). This model has been based on the principle that interferometric observations should be, at first, carefully validated for the quality of the survey. Secondly, we use the fact that the input catalog is larger than the final list of targets, so that we can optimize the night scheduling in terms of range of declination and hour angle during a single night, in order to support the stability of the transfer function and of the efficiency of the observations. The night scheduling is based also on an algorithm of priority, analyzing if a star appears in different programs, or if its observation has already been started. This tool will also be used to follow the coverage of the HR diagram and to take the necessary decisions if necessary.

The model is build around a database of observations which we have connected to the different classical tools developed on one side by the JMMC\footnote[4]{JMMC:~\url{http://www.jmmc.fr/}}, the french  center for optical interferometry, and on the other side by the SPICA group. A schematic description of this work is presented in Fig.~\ref{fig:spicadb}.

Based on the SPICA catalog, the Night Scheduling Software (NSS) is a graphical interface which allows the user to select a list of targets based on their observability in order to prepare an observing run. This selection of stars can be done using several criteria, such as the program names, the instrumental modes (uniform or limb-darkened diameter, imaging, or spectral resolution) and the CHARA instrumental set-up. Other filtering options associated to given instrumental modes are also available, such as the date of the observation or the required resolution. The user can also select stars based on a criteria indicating their priority to be observed. The NSS offers as well the possibility to make a selection based on the declination and the magnitude. It also allows the user to choose calibrators (whose diameters are taken from the JSDC catalog\footnote[5]{JSDC:~\url{https://www.jmmc.fr/jsdc}}). Once this selection done, it can be sent to the ASPRO2\footnote[6]{ASPRO2:~\url{https://www.jmmc.fr/aspro2}} software through the SAMP interoperability protocol as a VO table in order to finalize the preparation of the observing run. The UV coverages and models are available as well based on the SPICA catalog and can be retrieved using TAP/ADQL queries to follow VO concepts\footnote[7]{\url{https://www.ivoa.net/deployers/intro_to_vo_concepts.html}}.

\begin{figure} [htbp]
\begin{center}
\includegraphics[width=0.8\textwidth]{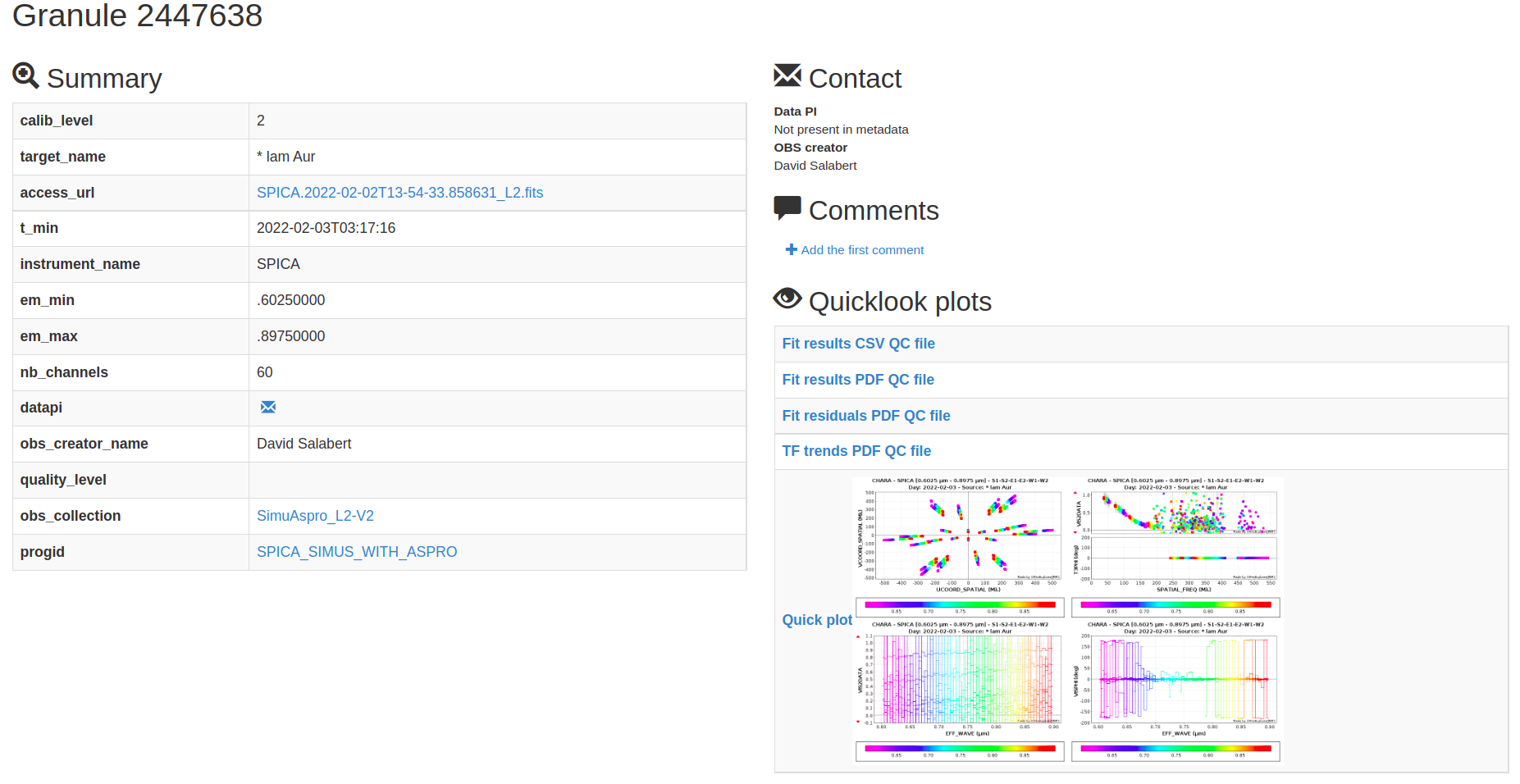}
\end{center}
\caption{Example of the SPICA database seen through the OiDB interface in the case of a simulated target.}
\label{fig:oidb} 
\end{figure} 

At the end of the night, once the raw data are reduced into L1 data by the Data Reduction Software (DRS), the Quality Control Software (QCS) 
computes the transfer functions and calibrates the L1 data into L2 data. It performs several tests to provide quality flags measuring the variability of the observed targets and updates the associated entries of the SPICA catalog. Data and ancillary files (transfer functions, quality checks, etc.) are then automatically submitted to the OiDB\footnote[8]{OiDB:~\url{http://oidb.jmmc.fr/index.html}} and ObsPortal\footnote[9]{ObsPortal:~\url{http://obs.jmmc.fr/}} databases (Fig.~\ref{fig:oidb}).\\

Finally, the follow-up of the SPICA survey is done using the Data Feeding Software (DFS). The DFS checks for previous observations already in OiDB and allows each PI to consult and see the progression of his/her scientific program. The PIs can then check the quality of the new observations and the results from the QCS on OiDB, modify and update relevant fields from the SPICA catalog (e.g., the priority), and validate the observations or not. 

All the developments linked to the SPICA database make an important use of the interoperability with the available JMMC softwares not only for the preparation of observations but also for the interpretation of the measurements\footnote[1]{\url{https://amhra.oca.eu/AMHRA/index.htm}}.

\section{First results of the commissioning of the instrument}

\subsection{Level of injection}
\begin{figure} [htbp]
\begin{center}
\includegraphics[width=0.8\textwidth]{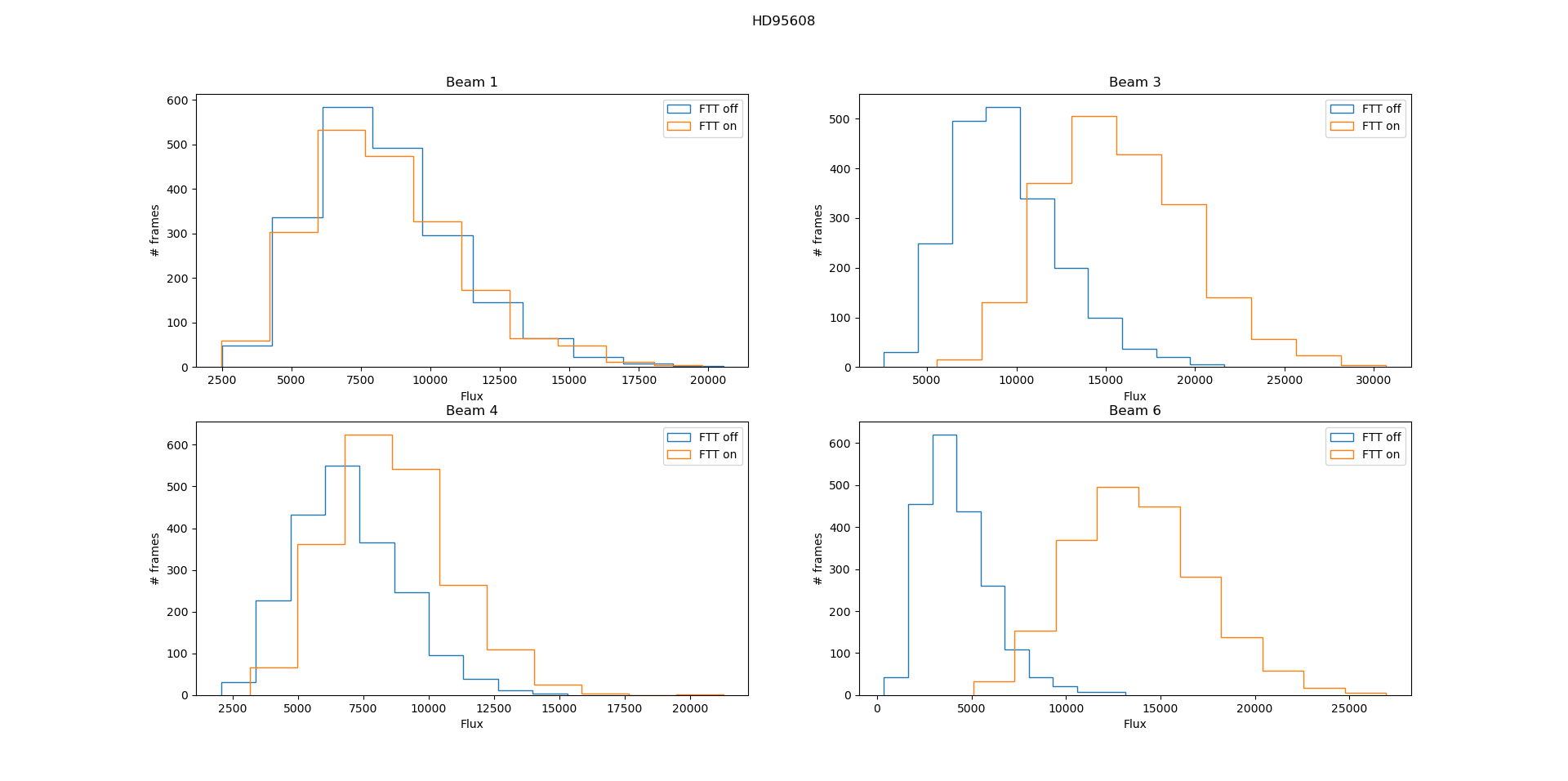}
\end{center}
\caption{Observation of HD95608 ($m_V=4.4$): Histogram of the flux injected into the fibers (spectral band 645-900nm). Beams 2 (W2 without Adaptive optics) and 5 (S1 not available at the time of these observations) were closed.}
\label{fig:spicahisto} 
\end{figure}

One of the most critical aspects of the commissioning of SPICA has been to characterize the quality of the injection of the light into the single mode fibers. Knowing that SPICA is working behind an adaptive optics system not able to produce high Strehl ratio\cite{narsireddy_adaptive_2020} in the visible, we decided\cite{spie2018} to help the injection by the addition of a fast tip/tilt mirror. During the commissioning we demonstrated that these tip/tilt stages are able to work at a frequency of 100Hz (currently limited by the USB readout mode of the ANDOR detector) up to magnitude 8. After these first tests, efforts have enabled a Camera Link readout mode and to increase the preamplifier gain of the detector to reach a frequency of 200Hz at magnitude 8. We observed different stars in different conditions for this purpose. In Fig.~\ref{fig:spicahisto} we present the histograms of the flux with (red curves) or without (blue curves) the operation of the fast tip/tilt. The difference of improvement on the different beams is interpreted as a current limitation due to the static aberrations of some of the CHARA telescopes. A global effort is thus considered in the coming months to improve the image quality of the beams on one hand and to optimize the operation of the adaptive optics on the other hand.

From measurements on the Six-Telescope-Simulator of CHARA (STS), we have demonstrated that the injection modules offer a transmission level between 35 and 50\% after a fine collimation of the beams. The difference among the beams is easily explained by static aberrations in some of the beams of the STS. This level of 50\% is quite close from what has been expected after tests in laboratory. For improving this, we consider now the addition of a motorized translation stage for the perfect longitudinal positioning of the single mode fiber, as it appears that an adaptation of the focalization between the sky and the STS could be necessary, at a level not easily seen on the STS. 

\subsection{Instrumental visibility and birefringence}
\begin{figure} [htbp]
\begin{center}
\includegraphics[width=0.8\textwidth]{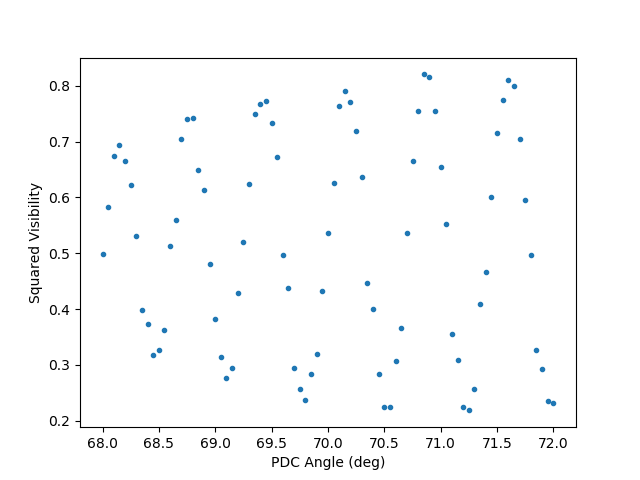}
\end{center}
\caption{Evolution of the visibility of the baseline B1B2 when changing the position angle of the PDC on B2, while keeping the PDC on B1 fixed. The modulation is due to the change in the delay in one polarisation on B2. The maximum of visibility is obtained around $70.8^{\circ}$.}
\label{fig:pdc} 
\end{figure}

As explained in Sec.~2.2, it is necessary to correct the difference of birefringence of the single mode fibers to optimize the instrumental visibility of SPICA. This process is done on the internal white light source of the STS, by pair of beams. The PDC module (Polarization Difference Compensator) on one beam is fixed, while the one on the other beams is slowly rotated to change the thickness of glass seen by the beam. This generates some 
changes in the optical path, different in the two polarizations. The resulting fringes exhibit therefore a variation of the instrumental visibility as a function of the position angle of the movable PDC. An example is presented in Fig.~\ref{fig:pdc}. 

Thanks to these measurements, we have been able to set all the PDC to their best positions and to optimize the different instrumental visibilities, ranging now from 0.6 to 0.85. These values may still be optimized on sky. They could also depend on the exact spatial sampling of each individual system of fringes.

\subsection{First fringes}
During a few nights in June 2022, first SPICA-VIS fringes were found on different nights. The six telescopes of CHARA were maintained in coherence thanks to the MIRC-X group delay tracking facility, using the new fast communication channel with the optical path length equalizers (see Anugu et al., these proceedings). The initial setting of the SPICA-VIS internal delay lines was set on the STS fringes. Unfortunately, the optical path in the Visible and in the IR is different from going from STS to sky. Therefore it was necessary to scan, in low resolution mode for SPICA-VIS, by steps of 20~$\mu m$, to find the visible fringes. S2S1 fringes (Beams 4 ad 5) were found at $\approx 300\mu m$ from the IR position, whereas the W2W1 (Beams 2 and 3) fringes were found at $\approx 3800 \mu m$.

For this initial operation, we were using the main longitudinal dispersion compensation of CHARA, but the visible longitudinal dispersion compensation optics\cite{Pannetier2021} were not active. They were just set at the position of their minimal thickness, this one having been determined thanks to the minimization of the dispersion of the STS fringes.

\begin{figure} [htbp]
\begin{subfigure}[b]{0.3\linewidth} \centering
\includegraphics[width=0.6\textwidth]{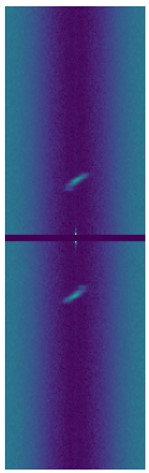}
\end{subfigure} \hfill
\begin{subfigure}[b]{0.3\linewidth} \centering
\includegraphics[width=0.6\textwidth]{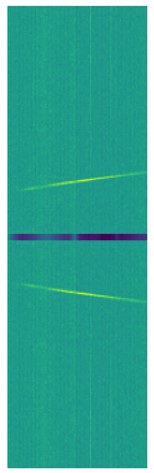}
\end{subfigure}
\begin{subfigure}[b]{0.3\linewidth} \centering
\includegraphics[width=\textwidth]{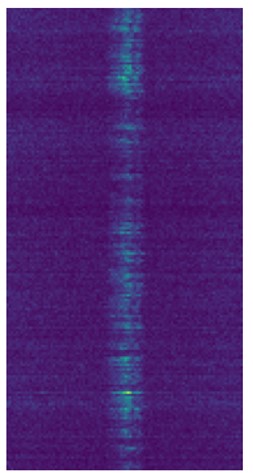}
\end{subfigure}
\caption{Deneb Observation: averaged 2D power spectrum of the interferograms (left) and averaged 1D power spectrum per spectral channel (Middle), Fringe waterfall (right). Fringes detected on the baseline S1S2.}
\label{fig:spicaDSP} 
\end{figure}

In Fig.~\ref{fig:spicaDSP}, we present a two-dimensional and a one-dimensional density power spectrum (DPS) of the fringes obtained on Deneb. The 1D-DPS presents the plot as a function of the spectral channel. In the same Figure, we present profiles of the DSP as a function of time, the so-called waterfall plot. This plot shows the stabilisation of the fringes and their detection on each individual frames.

In the coming months the SPICA and CHARA teams will be engaged to the continuation of this commissioning with the goal of demonstrating the science capabilities of SPICA-VIS and SPICA-FT on sky during the second 2022 CHARA semester. 

\section{Conclusion}
We have presented the main characteristics of the SPICA-VIS instrument installed at the focus of the CHARA Array. It is assisted by the SPICA-FT fringe tracker presented by Pannetier et al., this conference. The performance of the instrument is under investigation, and the first fringes having been found on sky just recently.

\acknowledgments{The development of SPICA (SPICA-VIS and SPICA-FT) has been made possible thanks to the funding support of CNRS, Université Côte d'Azur, Observatoire de la Côte d'Azur, University of Aarhus, Région Sud, and the CHARA Array. This program has made use of the Jean-Marie Mariotti Center services. This research has made use of the VizieR catalogue access tool, CDS, Strasbourg, France. This project has received funding from the European Research Council (ERC) under the 
European Union’s Horizon 2020 
research and innovation programme (Grant agreement No. 101019653). This work is based upon observations obtained with the Georgia State University Center for High Angular Resolution Astronomy Array at Mount Wilson Observatory.  The CHARA Array is supported by the National Science Foundation under Grant No. AST-1636624 and AST-2034336.  Institutional support has been provided from the GSU College of Arts and Sciences and the GSU Office of the Vice President for Research and Economic Development.}

\bibliography{spica} 
\bibliographystyle{spiebib} 

\end{document}